# The He I and He II chromospheric shells and the Transition Region

Cyril Bazin [1], Serge Koutchmy [1] and Ehsan Tavabi [2]

[1] Institut d'Astrophysique de Paris, UMR7095 CNRS and UPMC (France)
bazin.cyrille@neuf.fr ; koutchmy@iap.fr

[2] Payame Noor University, Zanjan, Iran
etavabi@yahoo.com

**Abstract.** Total eclipse observations were performed in 2008 and 2009 to study the He I and He II shells near the 1 Mm heights above the solar limb. They suggest that the corona penetrates deep into the chromosphere following magnetic chanels. Thanks to the use of a fast CCD camera, the observation of a second ionized helium shell is evidenced for the first time. The transition region is then seen at very low altitude where spicules are emerging. Spicule feet are also discussed, using the best resolution SOT/Hinode HCaII images processed with the non linear operator Madmax to look at details of this ubiquitous part of the solar atmosphere.

## 1 Introduction

New results coming from the August 1st 2008 and July 22nd 2009 total solar eclipses observed in Siberia and China were obtained in the frame of 2 expeditions leaded by L. Damé (CNRS). In both cases, a small experiment with a 600 grooves/mm transmission grating put in front of a 600 mm focal length refractor was prepared. In the first part of this work, we will make assumptions of an uniform layer with hydrostatic stratified shells, described for ex. by the VAL model Vernazza, Avrett and Loeser, 1981 [1]. We then measure the thickness and intensities profiles of the helium lines and of lines from heavier elements such as iron, ionized barium, titanium, etc. After, we consider the non homogenous nature of this layer.

## 2 Origin of the He II line : historical notes

The solar origin of this line in deep layers is badly understood. The ionization energy of He II is 54 eV. This line also corresponds to more energetic levels than the well known He II Lyman alpha line at 304 A. The temperature of formation is believed to be around or more than 50000 K ,Mariska et al 1992 [2]. This high temperature is possible in case of collisional effects due to beams of fast electrons coming from the hot corona or more probably from soft coronal radiations illuminating the chromosphere Zirin, 1975 [3], Avrett and Koutchmy 1989 [4], Mauas et al 2005 [5]. We do not know exactly how the fast electrons are guided or confined by the magnetic fields and how the helium lines are ionised in the low chromosphère, but some theoretical modelisations were done by Athay 1965 [6], [7]. These magnetic fields are concentrated by the converging motion of convective cells at the supergranulation boundaries Baudin et al 1997 [8]. Those cells produce the expulsion of magnetic fluxes with small-scale eruptions, such as spicules, jets, etc Filippov et al 2000 [9]. Another more plausible explanation could imply the photoionisation by X and EUV radiations coming from the surrounding hot corona, in case they penetrate deep enough into the 2 Mm thick layer, Avrett. and Koutchmy 1989 [10], Auchere et al 2000 [11]. In first approximation, the chromospheric layers as described by the VAL hydrostatic model with uniform stratified layers are where the phenomena occur. The Pashen $\alpha$ line of the He II line at 4686 A is known for a long time Butler 1925 [12]. It is observed in solar prominences, in galactic active nuclei, supernovae, in tormented Wolf-Rayet Stars, accretion disks, mini black-holes etc…It is very important to notice that in case of solar observations, only eclipses are free of parasitic light and can show the deep layers of the solar atmosphere where this of 16 km/frame on the limb at the August 1st 2008 total eclipse, and a linear CCD detector is used. line is produced: thanks to the occultation occuring in space, no parasitic scattered light is present like in out of eclipse observations, Pierce 1968 [13]. It is then possible to study the faint shells up to the coronal level intensities. Worden et al. 1973 [14] did not succeed in observing the HeII line at the limb with the SPO excellent telescope and spectrograph outside an eclipse. For our eclipse study, we used a rather small grating-objective lens, fitted with a fast CCD camera working at 25 frames/s to observe the helium lines in the region of 470 nm, during the second and the third contact of the eclipse. The temporal resolution is much higher than what has been done before, see Hirayama and Irie 1984 [15]; we obtained a resulting resolution

## 3 Eclipse observations of 2008 and 2009

The recorded spectra are over-exposed at the beginning of the 2nd contact, and the flux decreases very rapidly. A few seconds after, the emission lines appear, with the continuum of the solar limb between myriads of faint emission lines. The line intensities are modulated by the valleys and mountains on the limb profile of the Moon. The following images show a sample of flash spectrum fitting exactly with the Moon profile (courtesy of P. Rocher from the IMCCE France)

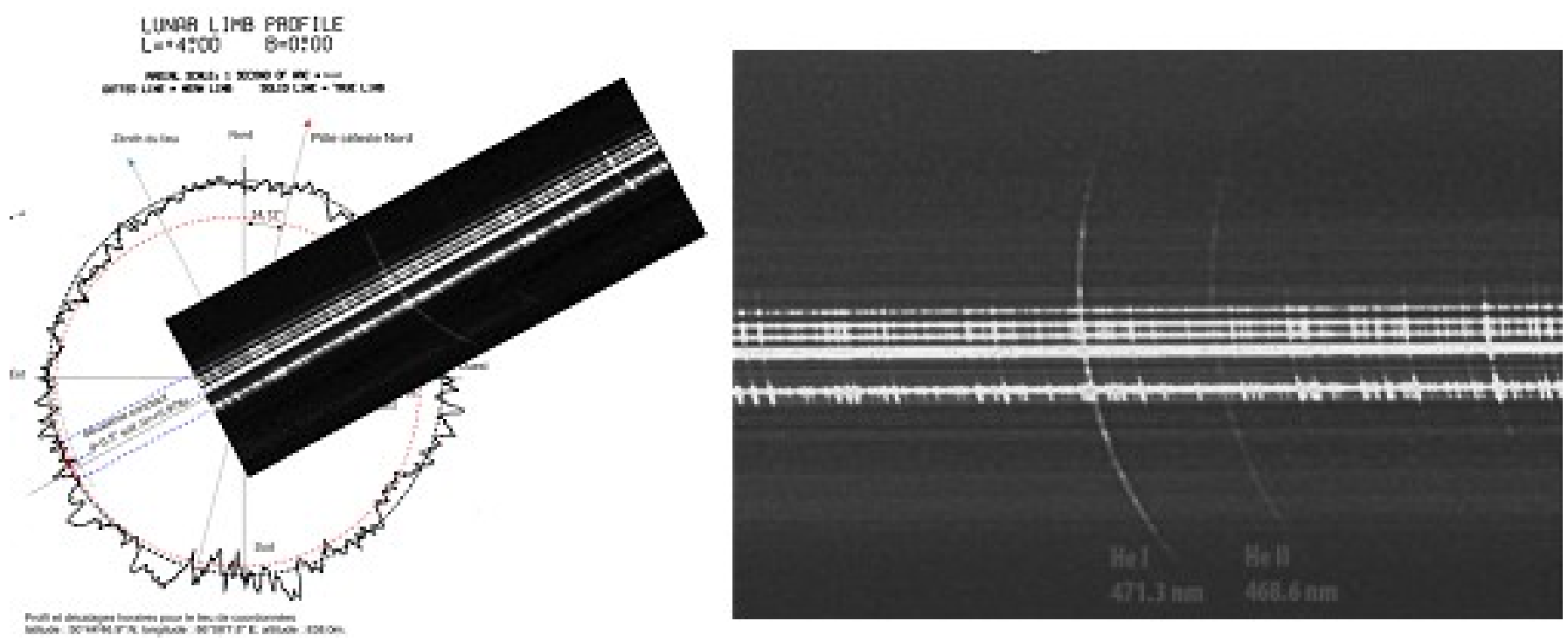

**Fig. 1.** Part of the flash spectrum coming from the light recorded above the lunar limb (at left). During the 2nd contact, about 250 spectra were obtained during the occultation process. Extract of the sequence obtained during the 2nd contact at the August, 1st 2008 eclipse (at right). Spectral resolution is 0.24 A/pixel. Spatial resolution across the dispersion is 1.8 Mm/pixel. Average of 10 spectra. Cadence is 25 frames/s.

The CCD camera allows obtaining a spatial resolution of 2 arcsecond/pixel. As said before, the advantage of the eclipses is that the phenomenon occurs in space. The occultation process is not affected by the Earth atmospheric turbulence. A greater extend of the chromosphere around the Sun appears as shells that entirely surround the Sun, in the line of the neutral helium (471,3 nm), and also of the ionized helium (468,6 nm). Figure 1 shows these shells in emission, in addition to the fainter low excitation emission lines of Fe II, Ti II, Mg I etc...
The details of the lunar limb (mountains and valleys) strongly modulate the intensity of the spectrum of the thin layers of helium shells. The intensities of the ionized helium line vary in a rather monotonous way over the crescent, because of the difference in relative curves between the Moon and the Sun. The line intensities were measured in successive frames in order to analyse the flux variations with the heights above the photosphere. The measurements were done in the lunar valleys, where the Bailey's bead spectra are seen, see Bazin and Koutchmy, 2009 [16]. By averaging the measurements, we obtained the curves (figures 2 and 5) of the line emission variations, and performed a comparison with the continuum variations, see also Kurokawa, 1974 [17]. The effective thickness of the ionized He II shell is estimated to be approximately 1500 km.

## 4 Analysis of the variations in the He I and He II lines

Figure 2 shows the intensity profiles of the two helium lines. The continuum is measured in the same region and the intensity unit is the "limb" intensity at a distance of approximately 20" of the edge of the Sun.

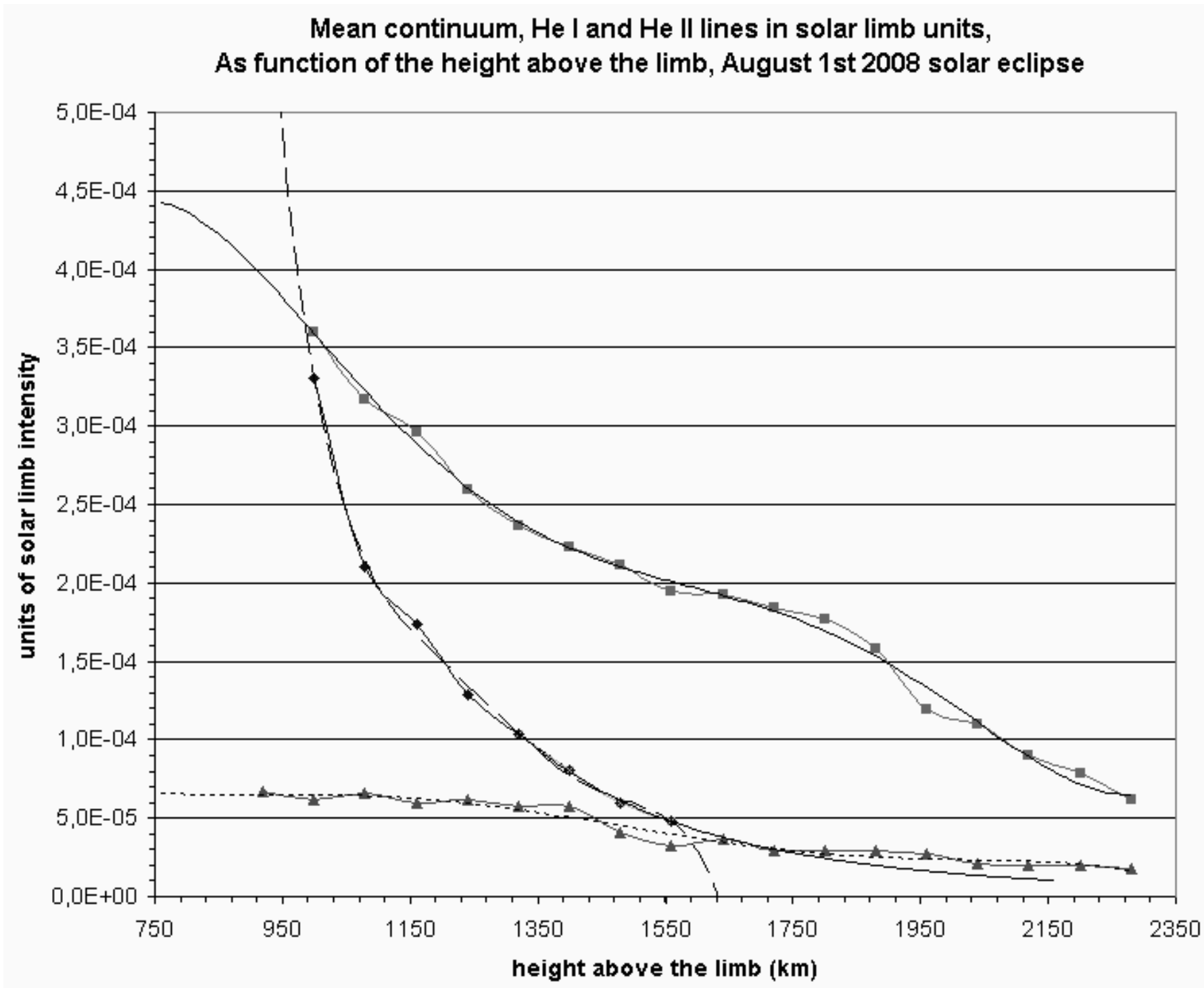

**Fig. 2.** Intensity profiles of the helium lines and continuum. 10 flash spectra stacked every five spectra. The true continuum was measured between emission lines. From the flash spectra taken at the 2nd contact of the August 1st 2008 total eclipse in Siberia. Heights are given in relative units

These profiles were obtained by averaging the intensities in each image along the helium lines thin crescent at the location of the valleys of the Moon.
Furthermore, at the July, 22nd 2009 total eclipse, flash spectra were obtained during the 2nd and the 3rd contact. Many lines were observed and identified such as the lines of Ba II, Ti II, Fe II and the line of the neutral helium at 447,1 nm, see figure 3. We used the Dunn's Tables from the 1962 eclipse [18]

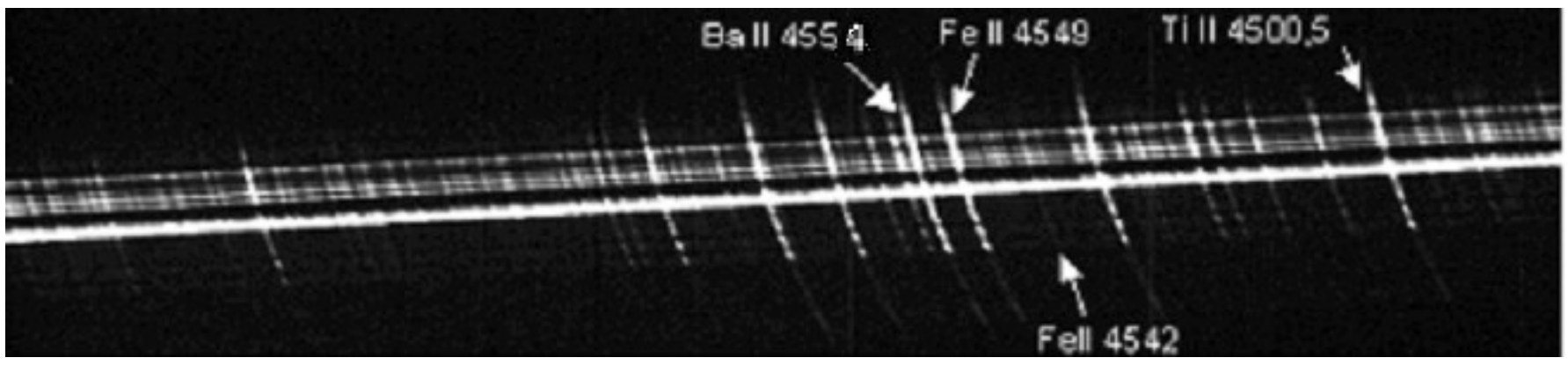

**Fig. 3.** 5 stacked flash spectra taken at the 2nd contact of the 22/07/2009 eclipse. Frame rate is 15 images/s with a CCD Skynyx 2.1 M Lumenera camera. Spectral resolution is .12 Angström/pixel. Spatial resolution on the azimuthal direction (normal to dispersion) is 0.9 Mm/pixel. Dunn et al. 1968 tables have been used for the lines identification.

Between the emission lines, the continuum can be accurately measured. Averages were possible thanks to the visibility of the Bailey's beads occurring near the contact. The intensities of the lines and continuum decrease and finally reach the coronal level.

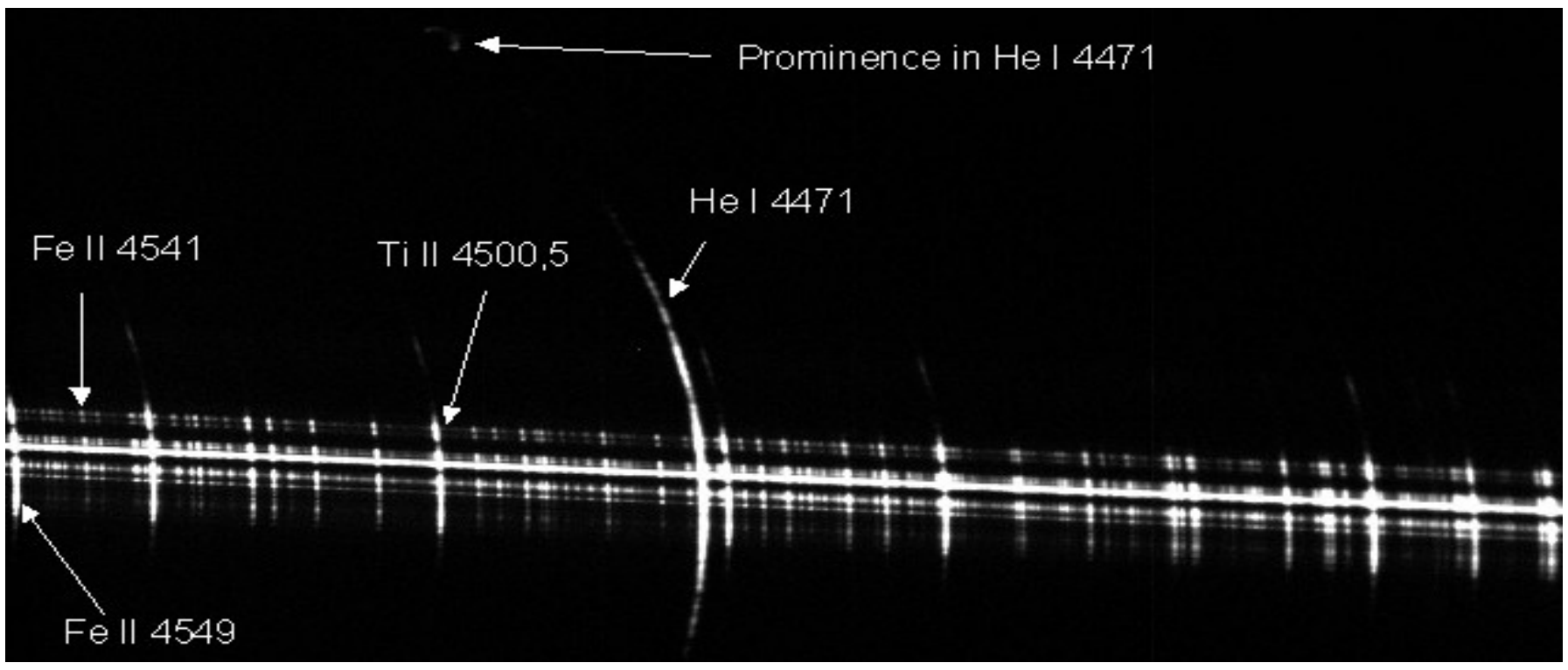

**Fig. 4.** *Sample of stacked flash spectra recorded at the 3rd contact showing the He I line at 447.1 nm*

Lines intensities were measured at the same position along the thin crescent. We obtained after averaging, the intensity profiles of the emission lines shown on figure 5.

It should be noted that these shells are also well observed on Trace images of the solar limb taken in the C IV UV line at 155.0 nm and also, very recently, with the SOT Hinode using the HCaII lines. A shell is clearly seen in emission with an approximately 2 Mm thickness corresponding to the ionised He II shell. Figures 6 show spicules and the location of their feet in the deep layers where the corona should penetrate.

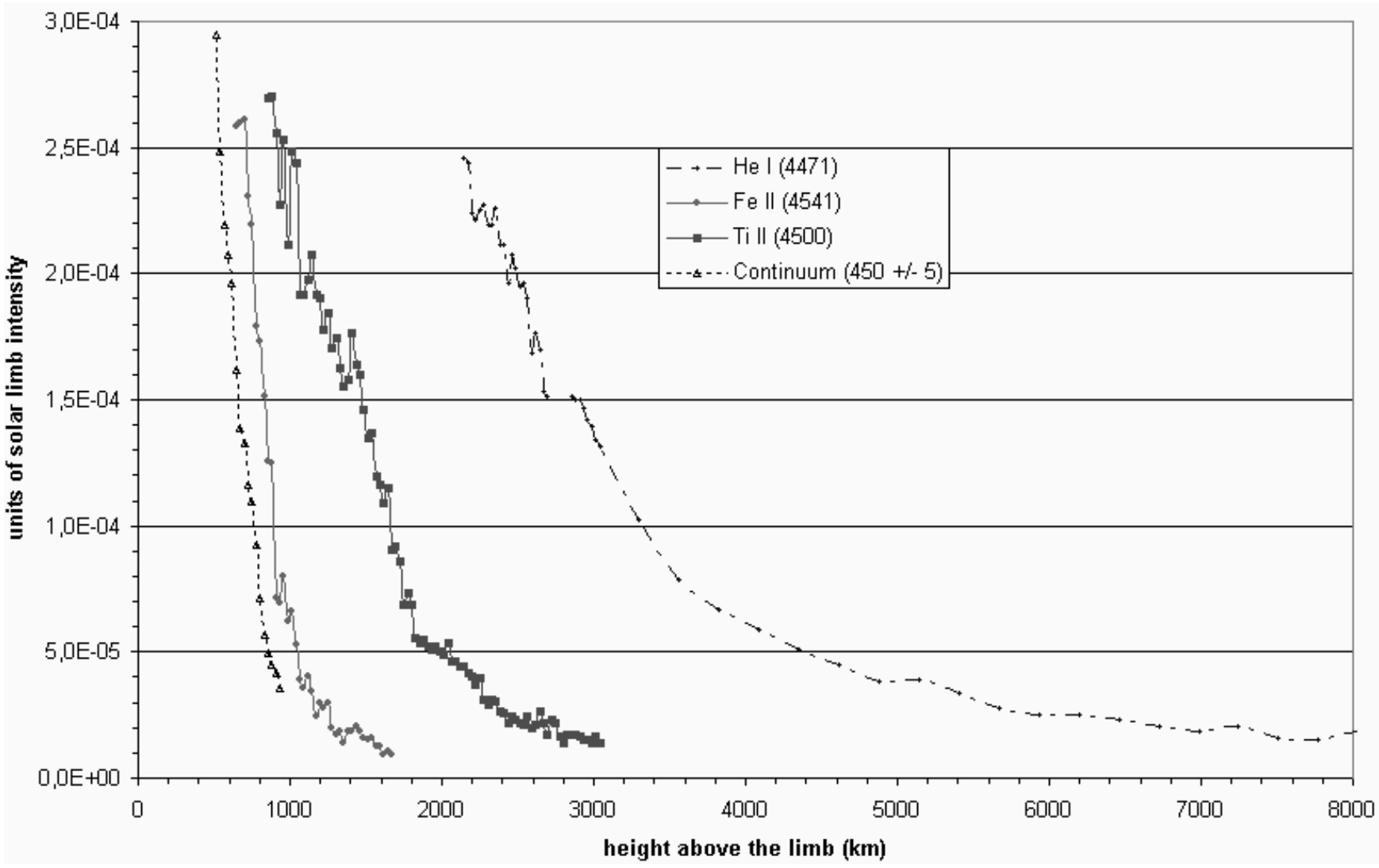

**Fig. 5.** Intensity profiles of He I, Fe II, Ti II and continuum deduced from the flash spectra recorded during the 3rd contact of the July, 22nd 2009 total eclipse, see figure 4

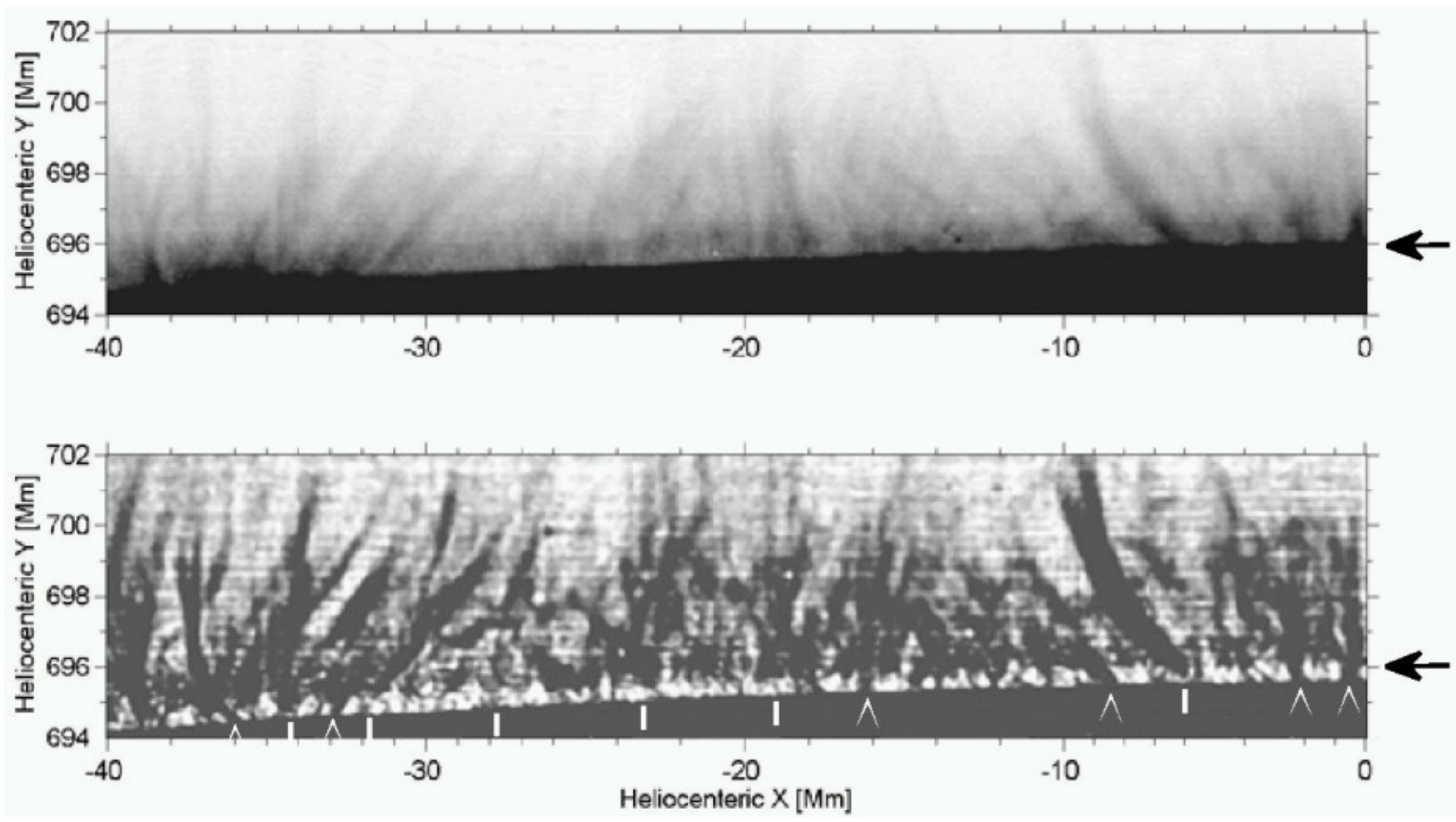

**Fig. 6.** HCaII SOT- Hinode limb original negative integrated for 5 min filtergram (upper part) and the corresponding madmaxed filtergram, see Koutchmy O. 1988 [19] reconstructed by superposing individual frames taken during 5 min at an 8 sec cadence. Note at the right of each image the arrows placed exactly at the same height to point out the "work" done with the madmax algorithm in showing very near limb structures. At the bottom of the madmaxed image, under the limb, some marks are put to help to recognize structures supposed to come from behind "Λ" or from beyond "l" the opaque photospheric limb.

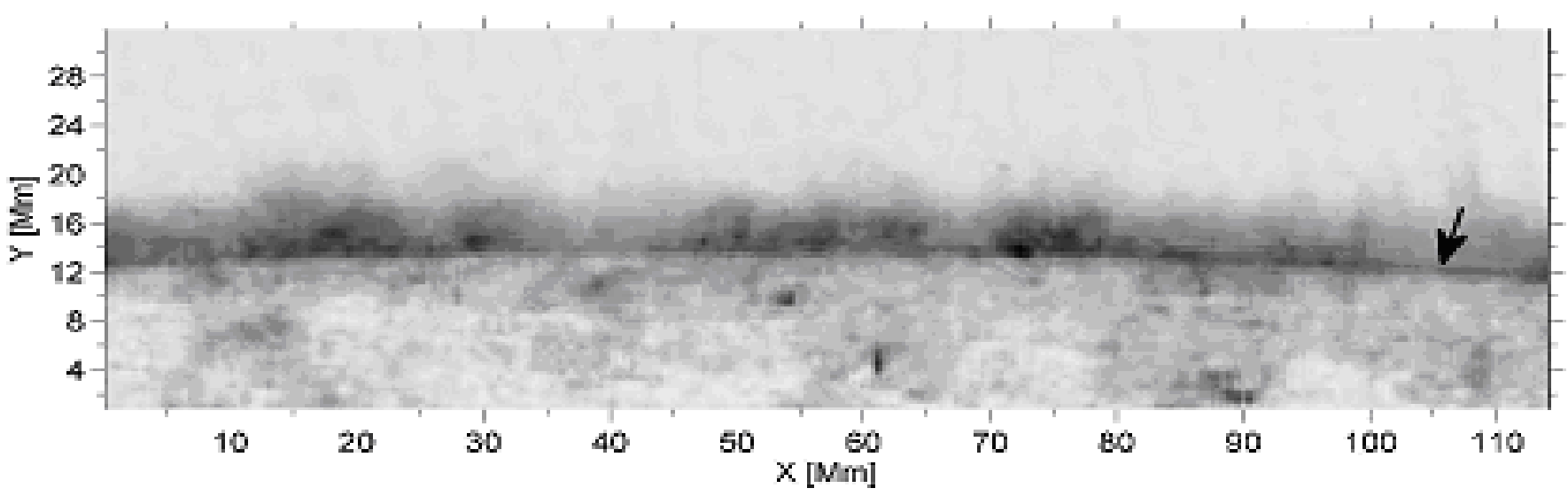

**Fig. 7.** Trace negative image of the polar limb (extracted from a sequence) obtained after subtracting 2 frames obtained at 1550 and at 1700 in order to try to show the CIV TR emissions only. No image processing was done after the substraction. The arrow points to a region where emissions are seen right above the limb with the thin shell well apparent.

The thin transition region is a rather dynamical phenomena. The magnetic field effects are responsible for the big jump of temperature (T< .01 MK to T > 1 MK) occurring above the limb. According to the VAL model, the transition region is put at the 2.3 Mm level. Our new results suggest that the TR where the He II line is produced should be lower than 2 Mm, see also Daw et al 1995 [20]. It was difficult to measure the fluxes very near the limb, under 1 Mm, due to the saturation of the signal and the limited dynamics of the used fast CCD[1]. The Madmax operator acts to substantially enhance the finest scale structure; it is a weakly nonlinear modification of a second derivative spatial filter. The Madmax processing on a 2D single average image uses the maximum of the second order derivative function of the intensities taken in 8 different directions, in order to increase the visibility of the features. This technique allows improving and revealing the hidden details on a picture, such as the feet of the spicules in the lower layers, and it helps us to measure and identify the heights above the limb where the ionized helium line should be created and the limits of the hot corona penetration.

## 5 Conclusions :

A deep helium shell is well identified at 468.6 nm as the Pashen $\alpha$ line of He II. The effective thickness of the shell is about 1500 km. The layer is assumed to be optically thin in this line.
The origin of the ionisation is not well understood. It could be by collision and/or by photoionisation and additional studies have to be done, including the study of the feet of spicules and their origin, see for ex. Athay 1956-1965 for the historical approach. This can be done in other spectral lines such as Ba II, Ti II, Mg II, and also with images from the Hinode and the Trace mission. New measurements of the Helium shells are planned for the next total eclipse of July 11$^{th}$ 2010.
The dynamic chromosphere needs to be studied in more details, such as the velocity, turbulence, Doppler shift of the spicules in HCaII, HeI and H$\alpha$ lines which allows recognising the feet of these structures in the deep layers, as illustrated by the Hinode filtergrams.
The continuum profile is well seen between emission lines. It can be used to improve the model of the temperature minimum region where p-modes (global oscillations) have a maximum of amplitude. To perform these studies, eclipses are the best way to analyse the faint helium lines, because the very low level of the parasitic scattered light is not possible to reach outside of eclipses.

[1] A last experiment using the fast 1K Lumenera CCD camera giving a true 12 bit depth of digitization was used at the total eclipse of July 11, 2010 in French Polynesia, with successful results which will be reported in a forthcoming paper.

**Acknowledgments.** We benefited from discussions with many scientists, among them Zadig Mouradian, Boris Filippov, Leon Golub, Luc Damé, that we sincerely thank.